\documentstyle[draft,epsf]{mn}

\def\gsim{\mathrel{\raise0.35ex\hbox{$\scriptstyle >$}\kern-0.6em 
\lower0.40ex\hbox{{$\scriptstyle \sim$}}}}
\def\lsim{\mathrel{\raise0.35ex\hbox{$\scriptstyle <$}\kern-0.6em 
\lower0.40ex\hbox{{$\scriptstyle \sim$}}}}
\def\oii{{\rm O{\sc ii}}}

\title{Testing the Hypothesis of the Morphological Transformation
from Field Spiral to Cluster S0}

\author[Kodama \& Smail]
{Tadayuki Kodama$^{1,2}$ \& Ian Smail$^1$ \\
$^1$ Department of Physics, University of Durham, South Road, 
Durham DH1 3LE, UK\\
$^2$ Department of Astronomy, University of Tokyo, Hongo,
Bunkyo-ku, Tokyo 113--0033, Japan}

\begin{document}

\maketitle

\date{Received, Accepted}

\begin{abstract}
{\it Hubble Space Telescope} observations of distant clusters have
suggested a steep increase in the proportion of S0 galaxies between
clusters at high redshifts and similar systems the present-day.  It has
been proposed that this increase results from the transformation of the
morphologies of accreted field galaxies from spirals to S0s.  We have
simulated the evolution of the morphological mix in clusters based on a
simple phenomenological model where the clusters accrete a mix of
galaxies from the surrounding field, the spiral galaxies are
transformed to S0s (through an unspecified process) and are added to
the existing cluster population.  We find that in order to reproduce
the apparently rapid increase in the ratio of S0 galaxies to
ellipticals in the clusters, our model requires that:  (1) galaxy
accretion rate has to be high (typically, more than half of the
present-day cluster population must have been accreted since $z\sim
0.5$), and (2) most of the accreted spirals, with morphological types
as late as Scdm, must have transformed to S0's. Although the latter
requirement may be difficult to meet, it is possible that such
bulge-weak spirals have already been `pre-processed' into the
bulge-strong galaxies prior to entering the cluster core and are
eventually transformed into S0's in the cluster environment.  On the
basis of the evolution of the general morphological mix in clusters our
model suggests that the process responsible for the morphological
transformation takes a relatively long time ($\sim 1$--3\,Gyr) after the
galaxy has entered the cluster environment.
\end{abstract}

\begin{keywords}
galaxies: elliptical and lenticular --- galaxies: formation ---
galaxies: evolution
\end{keywords}

\section{Introduction}

The morphologies of galaxies remain one of the key observational
characteristics used to classify and differentiate classes of galaxies.  It
is known from observations of the local Universe that galaxy morphology
depends on environment, with the galaxy population within clusters
dominated by early-types and spirals being found predominately in
lower-density environments.  Even within the clusters, the morphological
mix changes with local galaxy density (and hence with cluster-centric
distance) -- producing a morphology-density relation -- as shown by
Dressler (1980, D80).  Moreover, similar segregation behaviour has been
seen in {\it Hubble Space Telescope} ({\it HST}) imaging of ten
clusters at higher redshifts, $z=0.37$--0.56 by Dressler et al.\ (1997,
D97). These results confirm galaxy morphologies as one of the most sensitive
tracers of environmental influences currently known.

One of the most interesting results to come out of the morphological
studies of distant clusters has been the claim by Dressler et
al.\ (1997) that the S0-to-elliptical (S0/E) ratio is a strongly
decreasing function of redshift: by a factor of $\sim 5$ between $z\sim
0$ and 0.5 (although see Andreon 1998; Fabricant, Franx \& van Dokkum
2000).  This contrasts with the increasing number of blue, spiral
galaxies seen in these regions at higher redshifts; the origin
of the Butcher-Oemler effect (Butcher \& Oemler 1984; Dressler et
al.\ 1994; Couch et al.\ 1994, 1998; D97).  D97 therefore proposed that
most of the cluster S0's are formed relatively recently by the
morphological transformation of spirals (see also Poggianti et
al.\ 1999; Fabricant et al.\ 2000).  These spiral galaxies are
continuously supplied by accretion from the surrounding field in the
course of the assembly of the cluster.  Couch et al.\ (1998, C98) and
Fasano et al.\ (2000) have added estimates of the S0/E ratios for three
clusters at $z=0.31$ and nine clusters at $0.09<z<0.26$, respectively,
to bridge the gap in the D97 analysis between $z=0.37$ and the
present-day.  These studies confirm the trend of a decline in the
cluster S0 population from $z=0$ out to $z\gsim 0.5$.  Proposed
mechanisms for transforming the morphologies of bulge-strong spirals to
S0 galaxies include dynamical interactions, such as harassment (Moore
et al.\ 1996, 1999a), while gas dynamical processes, e.g.\ ram-pressure
stripping (Abadi, Moore \& Bower 1999), will also aid in truncating
star formation and hence fading the disk component of any spiral galaxy
within the cluster environment.

The luminous galaxy populations of rich clusters at $z\sim 0$ are
dominated by passive galaxies with early-type morphologies and
apparently old  stellar populations, as shown by their tight
colour-magnitude relation (e.g.\ Bower, Lucey \& Ellis 1992; Terlevich
et al.\ 1999). The transformation of star-forming spiral galaxy to
passive cluster member {\it must} occur, as the blue, spiral galaxies
we see in clusters at $z\sim 0.5$ cannot escape and hence their
descendents {\it have} to reside in these regions at the present-day
(Bower, Kodama \& Terlevich 1998; Smail et al.\ 1998; van Dokkum et
al.\ 1998).  Following the work by Smail et al.\ (1998) and Poggianti
et al.\ (1999) on the relationship between the blue galaxies in distant
clusters and nearby cluster S0's, Kodama \& Bower (2001) quantified the
analysis by sketching the flow of galaxies across the colour-magnitude
diagram down to the present-day, taking into account the galaxy
accretion from the surrounding fields.

Motivated by this success in connecting the photometric evolution of
the galaxy populations in distant clusters to the present-day
counterparts within the framework of the assembly of rich clusters, we
now try to seek to explain the evolution of the morphological mix in
clusters, and especially the decline in the S0 population, 
using a simple model for the growth of clusters and the evolution of
their galaxy populations.

Since the proportion of spiral to elliptical galaxies in the field is
much higher than that in clusters at any time from $z\sim 0.5$ down to
the present-day (e.g.\ Driver et al.\ 1998; van den Bergh et
al.\ 2000), continuous accretion of galaxies from the field should be
an effective process in changing the morphological mix in clusters.
If field spiral galaxies are turned into S0's upon accretion on to clusters,
then the  accretion process effectively increases the fraction of S0
galaxies with time, and might explain the rapid increase of S0/E ratio
presented in D97.  The aim of this Paper is to test this hypothesis
quantitatively and thereby constrain both the galaxy accretion rate
and the timescale of the morphological transformation.
Unless otherwise stated, we use H$_0=50$\,km\,s$^{-1}$\,Mpc$^{-1}$ and
q$_0=0.1$.

\section{The Morphological Mix in Distant Clusters and the Field}

\subsection{The Morphological Mix in Clusters}

To investigate the morphological evolution of the galaxy population in
clusters we use observations of galaxy morphologies in three samples of
rich clusters spanning $z=0$--0.5.  These come from the  survey of
local clusters by Dressler (1980), where we use the same ten high
concentration clusters as in D97,   and two different {\it HST}--based
studies: Couch et al.\ (1998)'s work on three clusters at $z=0.31$ and
the morphological catalogue of ten rich clusters at $z=0.37$--0.56 from
Smail et al.\ (1997b), which is the basis of the D97 analysis.

To reliably compare the galaxy samples at different redshifts we must
apply the same selection criteria to each survey.  Specifically, when
attempting to trace the evolution of galaxies between high redshifts
and the present day, we have to take into account the various degrees
of fading they will suffer after their star formation ceases, which
depends upon their star formation histories.  For example, the most
vigorously star forming late-type spirals in the distant cluster
samples may fade substantially ($\lsim 1$\,mag) after they cease star
formation (see Kodama \& Bower 2001) and therefore only the brighter
examples ($M_V\lsim -21$) will appear in a local cluster sample limited
at $M_V\sim-20$.  For the simple model discussed here we have chosen to
ignore the added complication of a star-formation dependent magnitude
cut and simply adopt a fixed magnitude limit for all morphological
types.  The D80 sample is limited at an absolute magnitude equivalent
to $M_V\sim -20.4$ (D97).  We have therefore also chosen  to  limit the
more distant samples from D97 and C98 at an absolute magnitude of
$M_V=-20$ (we note that the magnitude limit quoted in D97 for their
analysis is incorrect and actually corresponds to $M_V=-19$). By
adopting a fixed absolute magnitude cut we will tend to over-estimate
the proportion of descendents of late-type spirals predicted in the
local clusters.  However, we note that the fraction of late spirals
(Scdm) is small (7 per cent) and therefore our results do not
significantly change if we use, for example, a $M_V=-21$ cut for this
population.

In addition, because of the morphology--density relation (D97), it is
important to use equivalent areas in the nearby and distant clusters
when comparing the galaxy populations. The morphological mix in the
high redshift clusters in D97 and C98 is calculated within the {\it
HST} fields, which roughly correspond to the inner 130$''$ or 1.0\,Mpc
($z\sim 0.5$) diameter regions.  For the nearby sample from D80
therefore we use the same physical radius to define the morphological
sample.

We show in Figs.~1 and 2 the observed S0/E and Sab/S0 ratios,
respectively, from  D80, C98 and D97.  For the local sample of D80 we
plot both the individual clusters as well as the average and the
scatter for the whole sample.  We note that observations of all of the
ten clusters are consistent with the mean value given the large
uncertainties of the individual cluster measurement arising from
sampling errors.

\subsection{The Morphological Mix in the Field}

To successfully model the evolution of the morphological mix in
clusters we need to know the distribution of morphologies for the
galaxy population they accrete from the field.  Ideally the
morphological mix for this field sample would be based on  a large,
volume-limited sample culled from a morphologically-classified faint
field galaxy redshift survey.  Unfortunately, there are no large
redshift surveys with adequate morphological information and so we have
instead chosen to use a magnitude-limited morphological survey at a
depth which should place the majority of the galaxies within the
redshift range we require.  We therefore use the  morphological
classifications from the {\it HST}--based Medium Deep Survey (MDS,
Griffiths et al.\ 1994) which reaches $I=22$ and will have a median
redshift of $z \sim 0.5$ with 75 per cent of the galaxies at $z\leq
0.65$, based upon the redshift distribution at this depth in Lilly et
al.\ (1995).  This means that the majority of the sample represent the
population which will be accreted onto the clusters in the redshift
range covered by our analysis.  The morphological mix of field galaxies
for this sample is then:  E, $0.10\pm 0.01$; S0, $0.10\pm 0.01$; Sab,
$0.25\pm 0.01$; Scdm, $0.30\pm 0.01$; and Irr, $0.25\pm 0.01$ (as used
by Smail et al.\ 1997b).  The errors are based on Poisson statistics.
This distribution is consistent with the smaller sample of van den
Bergh et al.\ (2000) in the {\it Hubble Deep Field} with spectroscopic
redshifts which place them at $z=0.25$--0.6, as well as the small
number of field galaxies with morphological and spectroscopic
information from Dressler et al.\ (1999).  This then provides us with a
morphological distribution for the field galaxies which are accreted
onto clusters at $z<0.5$.

In a similar manner to that discussed above, when comparing the cluster
populations at different epochs, we also need to consider a variable
magnitude limit when estimating the properties of the field galaxies
which are accreted onto the clusters, depending upon their previous
star formation histories (or crudely the morphological type).  However,
the lack of any large morphologically-classified faint galaxy redshift
surveys also means that we do not have the information available to
reliably apply a variable apparent magnitude cut to the different
morphological subsamples in the field population.  To quantify the
likely error which will arise from adopting a single magnitude cut
across all morphological types we turn to the type-dependent luminosity
functions available for local galaxy samples\footnote{While there is no
suitable high redshift sample to use, we note that there is no evidence
for differences in the evolution of the bright end of the luminosity
functions for colour-selected (red and blue) galaxy populations out to
$z\sim0.5$, at least down to our limit of $M_V=-20$ (Lilly et
al.\ 1995).} from the 2dF Galaxy Redshift Survey (Folkes et
al.\ 1999).  Correcting the $B$-band luminosity functions to $V$-band
using for type-dependent $(B-V)$ colours from RC3 (Buta et al.\ 1994),
and adopting limits of $M_V=-20$ for field E/S0s and $M_V=-20$ or
$M_V=-21$ for spirals (Sab+Scdm) we can estimate the likely variation
in the fractions of E/S0, Sab and Scdm.  Using a limit of $M_V=-21$ for
the spiral galaxies, we estimate relative proportions of 1:0.90:0.95
for E:Sab:Scdm, compared to 1:1.25:1.5 for the $M_V=-20$ cut (we note
that this latter ratio is very similar to the value derived above for
the distant field, using $M_V=-20$).  We conclude therefore that by
adopting a fixed magnitude limit across all morphological types in the
field we increase the proportion of spiral galaxies in the field by
factor of $\sim 1.4$--1.5 for both early- and late-type systems.  We
will discuss the consequence of this effect in \S4.

\section{Modelling the Evolution of Galaxy Morphology in Clusters}

Using the morphological distributions for the cluster and field
populations derived in the previous section we devise a series of
models for the evolution of the morphological mix in clusters from
$z\sim 0.5$ to $z=0$.  These models are by necessity simple and ignore
many of the details of the dynamical processes which may be involved in
the morphological transformation, such as galaxy harassment (Moore et
al.\ 1996; 1999a) and/or ram-pressure stripping (Abadi et al.\ 1999).
However, the relative importance of these processes is still a matter
of debate (e.g.\ Poggianti et al.\ 1999) and the details of the
processes are highly uncertain. We therefore adopt a phenomenological
approach to test the basic foundations of the morphological
transformation hypothesis for the evolution of the cluster S0
population.

We compare the predictions of the various models for  the evolution of
the S0/E and Sab/S0 ratios within the clusters with the observations
from D80, C98 and D97.  As we show below, these comparisons constrain
both how large a fraction of the cluster population must be accreted
since $z\sim 0.5$ and how long the morphological transformation takes.

There are three key parameters in the models: (1) the total galaxy
accretion rate onto the cluster; (2) the range of spiral types
(e.g.\ Sab, Scdm) which transform into cluster S0's, and (3) the
timescale of this morphological transformation.  We note that the
accretion rate and morphological range have almost the same effect on
the morphological evolution in the clusters, since both determine how
many of the accreted field galaxies are turned into S0's.  In contrast
the transformation timescale controls how long the accreted spirals are
classified as having `spiral' morphology after entering the cluster.
The parameters for the models we consider are summarised in Table~1.

\begin{table}
\caption{Summary of the Models}
\label{tab:models}
\begin{center}
\begin{tabular}{lccc}
\hline\hline
\noalign{\smallskip}
Models  & $A_{\rm recent}^{\dag}$ & Morphological  & $\tau_{\rm trans}$ \\
        &             & Transformation       &  (Gyr) \\
\hline
\noalign{\smallskip}
(a) hi-all-2   &  high          & Sab+Scdm $\rightarrow$ S0 & 2 \\
(b) lo-all-2   &  low           & Sab+Scdm $\rightarrow$ S0 & 2 \\
(c) hi-early-2 &  high          & Sab      $\rightarrow$ S0 & 2 \\
(d) hi-all-3   &  high          & Sab+Scdm $\rightarrow$ S0 & 3 \\
(e) hi-all-1   &  high          & Sab+Scdm $\rightarrow$ S0 & 1 \\
(f) shi-all-1  &  super-high    & Sab+Scdm $\rightarrow$ S0 & 1 \\
\noalign{\smallskip}
\noalign{\hrule}
\noalign{\smallskip}
\end{tabular}\\
Note --- $^{\dag}$ The three different accretion rate models have
cumulative galaxy accretion fractions at $z=0.5$ and 0.3, 
($A_{\rm recent}(z=0.5)$, $A_{\rm recent}(z=0.3)$), of:
low, (0.35, 0.2);\\ high, (0.5, 0.3); and super-high, (0.7, 0.4).
\end{center}
\end{table}

The cumulative galaxy accretion rate onto a cluster, $A_{\rm cum}(t)$,
is normalised to unity at $z=0$ (equivalent to $t=t_{\rm univ}$, the age
of the present-day Universe --- 16.6\,Gyr in our cosmology) as:
\begin{equation}
A_{\rm cum}(t_{\rm univ}) = \int_{0}^{t_{\rm univ}} A(t) dt = 1,
\end{equation}
where $A(t)$ is the normalised accretion rate at a cosmic time $t$.
To quantify the amount of recent accretion we define $A_{\rm recent}(t)$,
the integrated accretion rate from time $t$ (or redshift $z$) to the
present-day ($t=t_{\rm univ}$, or $z=0$):
\begin{equation}
A_{\rm recent}(t)=A_{\rm cum}(t_{\rm univ})-A_{\rm cum}(t).
\end{equation}
The actual form for the accretion rate is taken from the Extended
Press-Schechter theory (Bower 1991), assuming a power spectrum with
$n=-1.5$.  Based upon the weak lensing estimates of the masses of the
clusters used in D97's analysis (Smail et al.\ 1997a) we adopt a mass
for the present-day cluster of $\sim 10^{15}$\,M$_{\odot}$, or
100\,M$^\ast$.  Since the galaxy accretion history is strongly
dependent on the threshold mass for the substructure within the
clusters (Bower 1991), we take the fiducial threshold mass of 10, 30
and 50\,M$^\ast$ for the low, high and super-high accretion rate
models, respectively.

For a particular accretion history, the evolution of the morphological mix
(fraction of E, S0 and Sab galaxies) in clusters is calculated as follows:
\begin{eqnarray}
f_{\rm E}(t) & = & \frac{\displaystyle 1}{\displaystyle A_{\rm cum}(t)} \{ A_{\rm cum}(t_1) \times f_{\rm E}(t_1) + \nonumber \\
 & & (A_{\rm cum}(t) -  A_{\rm cum}(t_1)) \times f_{\rm E}^{\rm field} \},
\end{eqnarray}
\begin{eqnarray}
f_{\rm S0}(t) & = & \frac{\displaystyle 1}{\displaystyle A_{\rm cum}(t)} \{ A_{\rm cum}(t_1) \times f_{\rm S0}(t_1) + \nonumber \\
 & & (A_{\rm cum}(t) -  A_{\rm cum}(t_1)) \times f_{\rm S0}^{\rm field} + \nonumber \\
 & & (A_{\rm cum}(t-\tau_{\rm trans}) -  A_{\rm cum}(t_1-\tau_{\rm trans})) \nonumber \\
 & & \times f_{\rm Sp\rightarrow S0}^{\rm field} \},
\end{eqnarray}
\begin{equation}
f_{\rm Sab}(t)=
\frac{\displaystyle 1}{\displaystyle A_{\rm cum}(t)} \{ (A_{\rm cum}(t)- A_{\rm cum}(t-\tau_{\rm trans})) \times f_{\rm Sp\rightarrow S0}^{\rm field} \},
\end{equation}
where $t_1$ is the cosmic time at $z=0.54$ (9.8\,Gyr in our cosmology)
the mean redshift of the four most distant clusters in the D97 sample,
the epoch at which we normalise the fraction of ellipticals and S0's
($f_{\rm E}(t_1)=0.40\pm 0.08$ and $f_{\rm S0}(t_1)=0.16\pm 0.04$, see
below).  $\tau_{\rm trans}$ is the timescale for the morphological
transformation from a spiral to S0 after accretion, and $f_{\rm
Sp\rightarrow S0}^{\rm field}$ is the fraction of field galaxies that
are eventually turned into cluster S0's.  Based on \S2.2
this is taken as:
\begin{equation}
f_{\rm Sp\rightarrow S0}^{\rm field} = \left\{
\begin{array}{lll}
f_{\rm Sab}^{\rm field} & =0.25\pm0.01 & {\rm Model~(c)},\\
f_{\rm Sab}^{\rm field}+f_{\rm Scdm}^{\rm field} & =0.55\pm0.02 & {\rm otherwise.}
\end{array}
\right.
\end{equation}

In the following we also assume that any bulge-strong galaxies observed
in the clusters are in the process of morphological transformation into
S0's and we ignore merging of galaxies, since it is unlikely to be
important for typical galaxies in these rich cluster environments at
$z\lsim 0.5$, where the relative velocities are high, $\gsim
1000$\,km\,s$^{-1}$.

To normalise the model for the evolution of the morphological mix
within  distant clusters we start from the morphologically-classified
galaxy population seen within clusters at the highest redshifts covered
by our analysis, $z\sim 0.5$.  For this we use the average of fractions
for the four most distant clusters in the Smail et al.\ (1997b) sample
used by D97:  Cl\,0412$-$65 ($z=0.51$), Cl\,1601+42 ($z=0.54$)
Cl\,0016+16 ($z=0.55$) and Cl\,0054$-$27 ($z=0.56$).  These are: E,
$0.40\pm 0.08$; S0, $0.16\pm 0.04$; Sab, $0.32\pm 0.10$; Scdm, $0.07\pm
0.06$; and $0.05\pm 0.03$ Irr/compact/unclassified.  The errors show
the scatter between the individual clusters.  The Poisson errors are
always small; about half of the errors shown.  These numbers should
represent the morphological mix for a `typical' $z\sim0.5$ cluster,
since the fractions are simply averaged across the four clusters rather
than weighted by richness.

\section{Results}

We plot in Figs.~1 and 2 the model S0/E and Sab/S0 ratios as a function
of redshift calculated from our model, normalised to the average
cluster population at $z=0.54$.  Fig.~1 shows that the evolution of the
S0/E ratio depends largely on the recent accretion rate, $A_{\rm
recent}$, (compare Models (a),(b) and (f)) and/or what proportion of the
accreted spiral population are turned into S0's (Models (a) and (c)),
and that it is relatively insensitive to the transformation timescale
,$\tau_{\rm trans}$ (Models (a),(d) and (e)).  To match the strong
evolution seen in the observations we must have both a high accretion
rate, typically $\gsim 50$ per cent of the cluster population added
since $z\sim 0.5$, and also transform the majority of the accreted
spirals into S0s (not only bulge-strong Sab galaxies, but also
bulge-weak Scdm's).  We discuss the possibilities for transforming
bulge-weak spirals into S0's in \S5.  Based upon Fig.~1 we conclude
that Models~(a), (d), (e) and (f) are preferred.  We note that if we
applied the type-dependent magnitude cut for the accreted field
galaxies to take into account the fading effect discussed above, the
necessary accretion rate would need to be even higher since there would
be fewer bright spirals available to be added from the field.  This
would strengthen the conclusions presented here.  This result confirms
quantitatively the earlier discussion by Fabricant et al.\ (2000), who
suggested that a high accretion rate of field galaxies was needed to
explain the strong evolution in the morphological mix in clusters
between $z=0.33$ and the present-day.

A further constraint on the models comes from the spiral fraction
(Fig.~2).  As our model assumes that most spiral galaxies observed in
the cluster cores are undergoing a morphological transformation into
S0's, the fraction of spiral galaxies in clusters can be used to
estimate the timescale of this process (Poggianti et al.\ 1999).
However, the spiral fraction also depends on the accretion rate and/or
the proportion of the total spiral population which can transform into
S0's. As the S0 fraction depends upon the same parameters, we can
reduce this dependence by normalising the Sab fraction using the S0
fraction.   The Sab/S0 ratio thus depends primarily  on the
morphological transformation timescale (Fig.~2) and can be used to constrain
this timescale.

We show in Fig.~2 the predicted Sab/S0 ratios for our models and
compare these to the observations.  To illustrate the dependence on the
transformation timescale we show three models: (d), (a) and (e) for
which $\tau_{\rm trans}=3$, 2, 1\,Gyrs respectively.  As Fig.~2 shows,
if $\tau_{\rm trans}$ is too short (1\,Gyr in Model~(e)), the model
struggles to retain enough accreted galaxies with spiral morphologies
in the clusters before they are transformed into S0's. On the other
hand, Model~(d) ($\tau_{\rm trans}=3$\,Gyr) tends to over-produce the
proportion of cluster spirals compared to the observations.   Equally,
adopting a very high accretion rate model, Model~(f), which is
consistent with the S0/E ratio evolution (Fig.~1), and $\tau_{\rm
trans}$ of 1\,Gyr tends to underestimate the Sab/S0 ratios (Fig.~2).
Taking into account the systematic uncertainty in the normalisation of the
models at $z\sim0.5$ using the fractions of E and S0 galaxies in the
four distant clusters (Fig.~1 and \S3), our model suggests that the
timescale for the morphological transformation should be in the range
$\tau_{\rm trans}\sim 1$--3\,Gyr.

\begin{figure}
\begin{center}
  \leavevmode
  \epsfxsize 1.0\hsize
  \epsffile{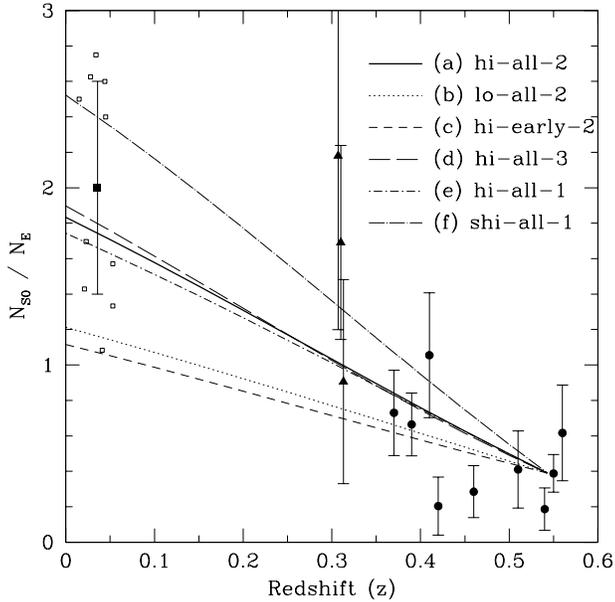}
\end{center}
\caption{
The predicted evolution of the S0/E ratio in the cluster cores.
The six lines show
Model~(a) [solid line]: high accretion, $t_{\rm trans}=2$\,Gyr,
Model~(b) [dotted line]: low accretion, $t_{\rm trans}=2$\,Gyr,
Model~(c) [dashed line]: high accretion, $t_{\rm trans}=2$\,Gyr,
Model~(d) [long dashed line]: high accretion, $t_{\rm trans}=3$\,Gyr,
Model~(e) [dot-dashed line]: high accretion, $t_{\rm trans}=1$\,Gyr,
and Model~(f) [long dot-dashed line]: super-high accretion,
$t_{\rm trans}=1$\,Gyr, respectively.
The morphological transformation of Sab+Scdm $\rightarrow$ S0 is assumed,
except for Model~(c) where Sab $\rightarrow$ S0 is assumed.
Observed ratios are taken from D97 (circles), C98
(triangles) and D80 (squares).
The average ratio in the D80 sample and scatter (1$\sigma$) are shown
by the filled square and the error-bar respectively.
Note, the systematic uncertainty in the normalisation of the model
based on the cluster populations at $z=0.54$ is $\pm$0.15 in the S0/E
ratio and this produces a systematic uncertainty in the S0/E ratios
at $z\sim 0$ of 15--20\%.
}
\label{fig:s0e}
\end{figure}

\begin{figure}
\begin{center}
  \leavevmode
  \epsfxsize 1.0\hsize
  \epsffile{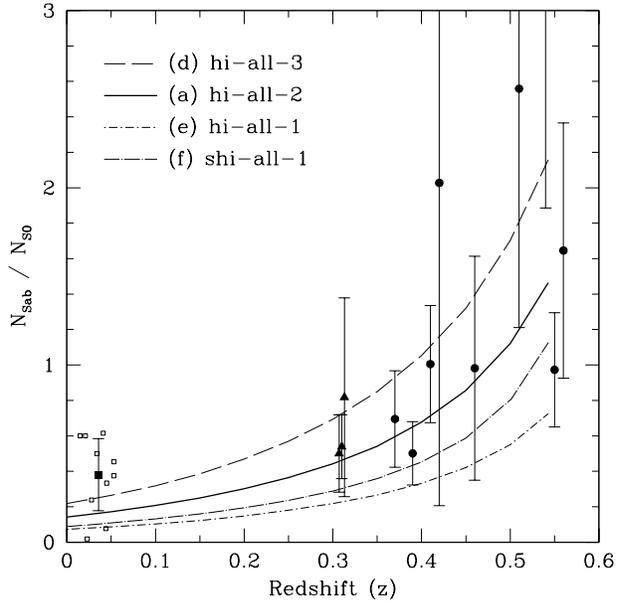}
\end{center}
\caption{
The evolution of the Sab/S0 ratio in our model clusters.  The four
lines show Model~(d) [long dashed line]: $t_{\rm trans}=3$\,Gyr,
Model~(a) [solid line]: $t_{\rm trans}=2$\,Gyr, Model~(e) [dot-dashed
line]: $t_{\rm trans}=1$\,Gyr, and Model~(f) [long dot-dashed line]:
$t_{\rm trans}=1$\,Gyr, respectively.  The high accretion rate is
assumed, except for Model~(f) where the super-high accretion rate is
assumed.  The data points are taken from D97 (circles) and C98
(triangles) and D80 (squares).  The average and scatter (1$\sigma$) are
shown by the filled square and the error-bar, respectively.
Note that the uncertainty in the model normalisation at $z=0.54$ using
the E and S0 fractions produces a systematic uncertainty in the Sab/S0
ratios of 30--35\% at $z=0.54$, 10--15\% at $z\sim0.4$, and $\sim$4\% at
$z\sim 0$.
}
\label{fig:ss0}
\end{figure}

Note that we have included field Scdm's into the cluster Sab category
in the models, except for Model~(c) (Eqs.~(5) and (6)), since it is
possible that the accreting Scdm's are transformed into Sab's either
prior to, or immediately after, their entry into the cluster cores
(\S5).  If we did {\it not} take into account this `pre-transformation'
from bulge-weak spirals to bulge-strong ones, the Sab/S0 ratios in the
models would go down by $\sim 50$ per cent.  In this case, only
Model~(d) would reproduce the observed trend, suggesting $\tau_{\rm
trans} \sim 3$\,Gyr.

The relatively long timescale required for the morphological
transformation in the model is consistent with a number of recent
results on the properties of galaxies in distant clusters.  Based upon
the paucity of post-starburst signatures in the spectra of the luminous
S0 population in distant clusters and the prevalence of red, passive
galaxies with late-type morphologies, Poggianti et al.\ (1999)
concluded that the morphological transformation of S0's had to occur on
a longer timescale than the stellar evolution lifetimes ($\gsim
1$\,Gyrs) of any A-stars formed in the most recent star formation event
in these galaxies (either while they were in the field, or any activity
which was induced on entering the cluster).  Comparisons between the
properties of the luminous S0 and elliptical galaxies in distant
clusters have been used to place limits on the relative evolution of
the two populations:  work on three clusters at $z=0.31$ -- using
spectral (Jones, Smail \& Couch 2000) and photometric (Couch et al.\ in
preparation) analysis concluded that the S0 and elliptical populations
are remarkably homogeneous, and similar conclusions were reached using
precise photometric analysis of the colours of S0/ellipticals in the
cluster cores at $z=0.18$ (Smail et al.\ 2001), $z=0.33$ (van Dokkum et
al.\ 1998) and  $z\sim 0.5$ (Ellis et al.\ 1997).  All of these studies
suggest that the luminous S0 galaxies in the cluster cores ceased their
star formation at least 2\,Gyr prior to the observed epoch (but see the
discussion in Smail et al.\ 2001 for less luminous S0's).  Observations
of the evolution of the colour distribution of cluster galaxies (Kodama
\& Bower 2001) and the radial gradients in star formation, measured
using [\oii], (Balogh, Navarro \& Morris 2000) both suggest that the
star formation in a galaxy declines relatively slowly after it has been
accreted by a cluster, $\sim 1$\,Gyr.

Taken together these results would indicate that an accreted
star-forming spiral galaxy transforms into a passive S0 within $\gsim
2$\,Gyrs of entering the cluster environment.  Our analysis confirms
this suggestions.  This is encouraging as our analysis relies purely
on modelling the evolution of the morphological mix within the
clusters, whereas the previous results are based on the stellar
populations within individual cluster galaxies and so these two
approaches are completely independent.

\section{Discussion and Conclusions}

We illustrate a simple phenomenological model to attempt to reproduce
the broad properties of the morphological evolution of cluster
populations across $\sim 7$\,Gyrs from $z\sim 0.5$ to the present-day.

In the framework of this model, we find that the observed evolution of
the S0/E ratio requires high rates of accretion onto the clusters and
also that a significant fraction of all accreted spiral galaxies, both
bulge-strong Sab and bulge-weak Scdm, are transformed into S0's.

The need to transform not only Sab, but also weaker-bulged galaxies
into S0's is a concern as there is no obvious and effective mechanism
for transforming an Scdm into an S0. The bulges of late-type spirals
are just not luminous enough to evolve into the bright bulges of S0
galaxies simply as a result of stopping the star formation in the disk
or even by `harassment' in the cluster environment (Moore et
al.\ 1996).  One possibility would be that the accreted field
populations are not the same as `pure' field population we assumed from
the MDS.  They may be `pre-processed' in group or sub-cluster units. or
intrinsically biased towards earlier types.  Most of the bulge-weak
Scdm's may have been already turned to bulge-strong galaxies by merging
or through bar-formation due to strong galaxy harassment/interaction,
before entering the clusters.  This is likely, given that the outskirts
of a rich cluster, where accreting spirals are located, also originate
from high density regions in the early Universe (e.g.\ Moore et
al.\ 1999b).  Therefore, the morphological mix in the field local to
the cluster may be skewed towards earlier Hubble types than that in the
`pure' field through galaxy--galaxy interactions in these denser
regions (Moore et al.\ 1999a, 1999b; Springel 1999; Diaferio et
al.\ 2001).

An alternative solution is to adopt a high density Universe.  With
$\Omega_o=1$ (q$_0=0.5$) for example  the accretion rate is higher than
in the current models  $A_{\rm recent}(z=0.5)\gg 0.7$ (Bower 1991). And
if this is really the case, the large S0/E ratio at $z\sim0$ can be
reached without the need to transform Scdm galaxies into S0's.
Furthermore, the S0/Sab ratio would also be consistent even with
$\tau_{\rm trans}=1$\,Gyr.  However, recent observations suggest slower
cluster evolution than expected in an $\Omega_o=1$ cosmology, with
little evolution seen in X-ray luminosity function of clusters out to
$z\sim 0.3$ (Ebeling et al.\ 1997) and beyond (Fairley et al.\ 2000),
indicating the recent accretion rate onto rich
clusters cannot be that high.

Finally, we note that there is the possibility of significant
scatter in the morphological mix in clusters even at the same
epoch (Figs.~1 and 2).  However, with the present observations
this scatter is consistent with the observational errors and so we have
chosen to model the evolution of the morphological mix in an
`average' cluster.  If real differences do exist between the
morphological distributions in similar clusters at a fixed 
redshift this would provide a powerful tool for investigating
the nature of galaxy morphology and the mechanisms responsible
for its transformation.

In summary, we have simulated the evolution of the morphological mix in
clusters with the  aim of applying a phenomenological test for the
hypothesis of the morphological transformation.  We therefore use a
simple model in which the accreted field spirals are transformed into
S0's in the cluster environment to attempt to match the observed
evolution of the distribution of morphologies for cluster galaxies out
to $z\sim 0.5$.  To reproduce the rapid increase in the S0/E ratio
between distant clusters and their present-day counterparts, it is
found that at least nearly half of the present-day cluster population
must have accreted since $z\sim 0.5$, consistent with the earlier
estimate by Fabricant et al.\ (2000).  Moreover, we found that for this
hypothesis to be correct, not only the bulge-strong field spirals, but
also relatively bulge-weak field spirals (Scdm) must have eventually
transformed into cluster S0's.  This  requirement may be difficult to
achieve unless the field population is pre-processed to produce
stronger-bulged galaxies prior to their arrival in the cluster core.
Finally, in order to explain the relatively high fraction of cluster
spirals we find that the process responsible for the morphological
transformation from spiral to S0s upon entry to clusters appears to be
relatively slow ($\sim$1--3\,Gyr).

\section*{Acknowledgements}

We thank Richard Bower, Michael Balogh, Warrick Couch, Alan Dressler
and Bianca Poggianti for useful discussions.  We also thank an
anonymous referee for their comments which significantly improved the
content and presentation of this paper.  TK thanks the Japan Society
for the Promotion of Science for support through its Research
Fellowships for Young Scientists.  IRS acknowledges support from the
Royal Society.

\end{document}